\documentclass{aa}
\usepackage{graphics}
\usepackage{astron}
\usepackage{amssymb}
\begin{document}

	\thesaurus{20(11.06.2; 11.07.1; 11.12.2; 11.16.1; 11.09.5; 11.19.6)}
%
	\title{Structure and stellar content of dwarf galaxies}
 	\thanks{Based on observations made at Observatoire de Haute
 	Provence (CNRS), France} \subtitle{${\mathrm{III}}$. B and R
 	photometry of dwarf galaxies in the M101 group and the nearby
 	field}

	\author{T. Bremnes \inst{1}
	\and B. Binggeli \inst{1}
	\and P. Prugniel \inst{2}}

	\institute{Astronomical Institute, University of Basel,
		Venusstrasse 7, CH-4102 Binningen, Switzerland
		\and Observatoire de Lyon, 
		F-64561 St.\ Genis-Laval Cedex, France}

	\offprints{T. Bremnes}
	\mail{bremnes@astro.unibas.ch}

\date{Received date / Accepted date}

\def\lacute{\mathopen{<}}
\def\racute{\mathclose{>}}

\maketitle

\begin{abstract}
We have carried out CCD photometry in the Cousins $B$ and $R$ bands of
21 dwarf galaxies in and around the \object{M101} group. Eleven are
members and suspected members of the group and ten are field galaxies
in the projected vicinity of the group. We have derived total
magnitudes, effective radii, effective surface brightnesses, as well
as galaxy diameters at various isophotal levels in both colours.
Best-fitting exponential parameters and colour gradients are also
given for these galaxies. Some of the galaxies show a pronounced
luminosity excess above the best-fitting exponential at large radii,
or surface brightnesses fainter than $\approx 
26 \mathrm{mag}/\sq{\arcsec}$ in $R$. This feature, while non-significant
for a single case and technically difficult to interpret, 
might be an indication of the existence of an extended 
old stellar halo in dwarf irregulars. The photometric parameters of the
galaxies presented here will be combined with previously published 
data for nearby dwarf galaxies and statistically analysed in a 
forthcoming paper.

\keywords{galaxies:general -- galaxies:fundamental
parameters -- galaxies:photometry -- gala\-xies:irregular --
gala\-xies:struc\-ture -- galaxies:luminosity function}
\end{abstract}

\section{Introduction}
A good knowledge of the local galaxian neighbourhood is prerequisite
for an understanding of the distant (early) universe. Deep images,
such as the spectacular Hubble Deep Field, can only be
interpreted properly if the dwarf galaxy content of the
local universe is very well known.  However, studies of nearby
dwarf galaxies have until recently concentrated on the Local Group
(LG), a rather small volume of space. A larger piece of the local
universe is captured in the ``10 Mpc Catalogue'' of galaxies by
Kraan-Korteweg \& Tammann \cite*{KKT79}, updated by Schmidt \& Boller
\cite*{1992AN....313..189S}.
This list is intended to contain all galaxies with radial velocities
of less than 500 km\,s$^{-1}$ as referred to the centroid of the LG,
i.e.~lying within a distance of about 10 Mpc. At present, the list
includes ca.~300 (mostly dwarf) galaxies, but this number is bound to
grow due to continued efforts to detect extremely faint and diffuse
nearby stellar systems \cite{1998A&AS..127..409K}.

Unfortunately, the available photometric data on ``10 Mpc objects'' is
relatively scarce and not very reliable for the fainter galaxies
\cite{patthuan96}. We have therefore started a long-term programme to
do systematic multicolour imaging of possibly all dwarfish objects in
the 10 Mp volume. The goal is not only to get total magnitudes in
order to assess the true shape of the faint end of the local
luminosity function of galaxies, but also to derive all relevant
structural parameters for these dwarfs and to compare them with
existing data on the dwarf galaxy populations of, foremost, the Virgo
and Fornax clusters e.g.\ Binggeli and Cameron
\cite*{1991A&A...252...27B}, in order to get clues on galaxy
evolution in different environnments.

Most galaxies within 10 Mpc distance are organized into a small number
of well-known groups of galaxies; essentially these are the IC342,
M81, M101, CVn\,I, Cen A, and Scl groups \cite{1992AN....313..329S}.
Following previous work on the M81 group dwarfs \cite{bre98,les99},
herefater Papers I \& II, we here present CCD photometric data in the
$B$ and $R$ photometric bands for the 11 known M101 group dwarf
members, as well as 10 field dwarfs in the vicinity of M101. A short
description of the M101 group is given in the following section.  The
photometric data presented here (Sect.~\ref{results}) will be lumped
together with those of previous data papers. An interpretation and
scientific discussion of this material is planned to follow in a
future paper of this series.

\section{Sample and imaging}
\label{sample}
The M101 group, with $D \approx$ 6.5 Mpc \cite{kar96}, is the most
distant one in the 10 Mpc volume. It is also the poorest group of all,
including the LG.  It is completely dominated by M101 itself: the
second-ranked group member, NGC\,5585, is already 3 magnitudes fainter
than M101. Only 13 members and possible members of the group are known
to date, half of which are lying very close to M101 and can therefore
be regarded as M101 satellites. With one exception (the dwarf
elliptical UGC\,8882) they are all late-type dwarfs (Sd, Sm, Im). One
peculiar feature of the group is its luminosity function: the
population of very faint and diffuse dawarfs (elliptical or
irregular), which is so frequent elsewhere, is apparently simply
missing here (the faintest member known is as bright as $M_{B}\sim
-14$). We have therefore made an attempt to find new candidate members
on deep POSS\,II Schmidt films, but found only one additional possible
member (the BCD MGC\,9-23-21). A recent blind H\,I survey of the M101
area has also not resulted in a single new member of the group
(Kraan-Korteweg et al.~1999, in preparation). It will take surveys of
highly increased sensitivity to uncover the sought-for exponential
rise of the luminosity function of the M101 group.

\begin{figure*}[htb]
\resizebox{12cm}{!}{\includegraphics{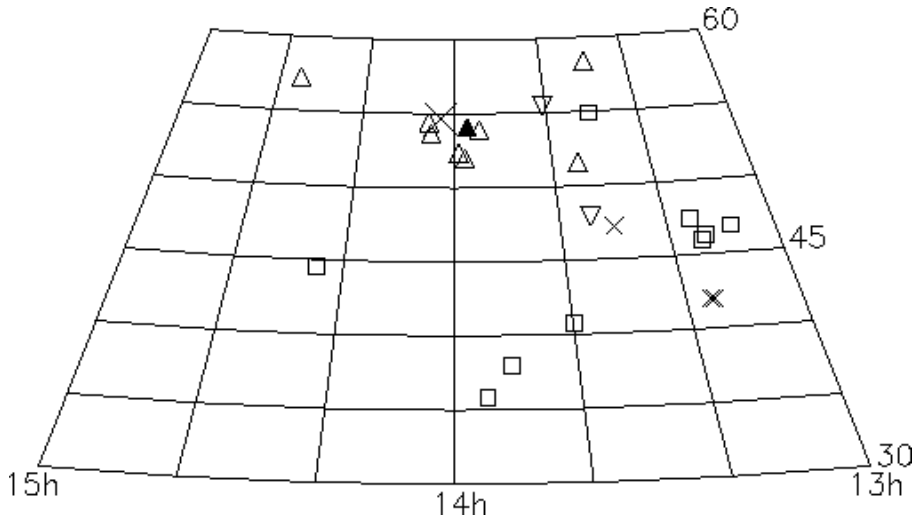}}
\hfill
\parbox[b]{55mm}{
\caption{Sky map (conical equidistant projection) of the M101 group
and surrounding field. The galaxies are coded according to group
membership and types. Group members and possible members are indicated
by triangles and inverted triangles, respectively, field galaxies by
squares. Early types are represented by filled symbols and late types
by open ones. M101 is plotted as a large cross. The galaxies M51 and
M63 are shown with smaller crosses. UGCA 342 is shown as a cross along
with M63 in the lower right of the map (see Sect.\ \ref{individual}).}
\label{map1}}
\end{figure*}

In Table~\ref{sampletable} we give a complete list of the 11 presently
known members (M) and possible members (PM) of the \object{M101} group
as well as 10 field (F) dwarfs that were imaged during the same run.
This list was prepared by B.\ Binggeli based on the catalogue of
Schmidt \& Boller \cite*{1992AN....313..189S}.  A map showing the
distribution of these objects on the sky is shown in Fig.~\ref{map1},
where the galaxies are coded according to their type and group
membership. A gallery of images is given in Fig.~\ref{images}. It
should be noted that the objects listed in Table \ref{sampletable} and
the images displayed in Fig. \ref{images} include {\em all}\/ M101
group members known to date with the exception of the two giant
members M101 and NGC\,5585 for which data and images are given in
Sandage and Tammann \cite*{1974ApJ...194..223S,1987rsac.book.....S}.

The photometry of the 21 objects listed in Table \ref{sampletable} is
based on images taken during eight nights in March 1997 on the
$1.2\:\mathrm{ m}$ telescope of the Observatoire de Haute-Provence
(OHP).  They are 40 minute Cousins $B$ and 20 minute Cou\-sins $R$
exposures. The camera used was the $\mathrm{ n}^{\circ}2$ Tektronix
$1024 \times 1024$ CCD. One pixel corresponds to $0\farcs69$, giving a
frame size of $11\farcm8 \times 11\farcm8$.  The gain was set to
$3.5\,\mathrm {e^-}$ per ADU, and the CCD was read out in the fast
mode, with a readout noise of $8.5 \,\mathrm {e^-}$. The seeing during
the observing run varied between $2.5\arcsec$ and $4\arcsec$ (FWHM),
which is relatively poor but sufficient for our purposes.

\begin{table*}
\caption[]{M101 members (M), possible members (PM) and Field dwarfs
(F) observed}
\begin{tabular}{crllcclccc}
\smallskip \\
\hline
\vspace{-3 mm} \\
Memb. &No. &Ident. 1 & Ident. 2 & R.A. & Dec. & Type & $D_{25}$ & $B_{\mathrm T}$ 
& $V_{\mathrm {hel}}$ \\
 & & & & (h\,\, m\,\, s) & (\degr\,\, \arcmin\,\, \arcsec) & & $(\arcmin)$ & (mag) & (km/s)\\
\vspace{-3 mm} \\
\hline
\vspace{-2 mm} \\
F & 1. & \object{UGC 08215} &\object{UGC 08215}	 &	13 08 03.3 	&46 49 43 &Im	&1.0 & 16.03    &218 \\
F & 2. & \object{DDO 167}   &\object{UGC 08308}  &	13 13 22.0	&46 19 07 &Im	&1.1 & 15.50    &164 \\
F & 3. & \object{DDO168}    &\object{UGC 08320}  &	13 14 26.1	&45 55 29 &Im	&3.6 & 13.04    &195 \\
  & 4. & \object{UGCA 342}  &\object{UGCA 342} 	 &	13 15 08.6	&42 00 10 &Im	&1.6 &          &388 \\
F & 5. & \object{DDO 169}   &\object{UGC 08331}	 &	13 15 30.7	&47 29 47 &Im	&2.7 & 14.27    &260 \\
M & 6. & \object{NGC 5204}  &\object{UGC 08490}	 &	13 29 36.4	&58 25 04 &Sm	&5.0 & 11.7{\textsuperscript{1}} &201 \\
F & 7. & \object{UGC 08508} &\object{UGC 08508}	 &	13 30 45.3	&54 54 34 &Im	&1.7 & 13.88    &62  \\
PM& 8. & \object{NGC 5229}  &\object{UGC 08550}	 &	13 34 02.8	&47 54 55 &Sd	&3.3 & 14.10    &364 \\
M & 9. & \object{NGC 5238}  &\object{UGC 08565}  &	13 34 42.8	&51 36 50 &Sdm	&1.7 & 13.55    &232 \\
F &10. & \object{DDO 181}   &\object{UGC 08651}  &	13 39 53.8	&40 44 21 &Im	&2.3 & 14.36    &201 \\
PM&11. & \object{UGC 08659} &\object{UGC 08659}	 &	13 40 33.9	&55 25 44 &Im	&1.0 & 16.16    &    \\
F &12. & \object{DDO 183}   &\object{UGC 08760}  &	13 50 51.1	&38 01 17 &Im	&2.2 & 14.64    &193 \\
F &13. & \object{UGC 08833} &\object{UGC 08833}  &	13 54 48.9	&35 50 17 &Im	&0.9 & 15.58    &228 \\
M &14. & \object{HO\,{\sc iv}}&\object{UGC 08837}&	13 54 45.1	&53 54 17 &Im	&4.3 & 13.65    &144 \\
M &15. & \object{UGC 08882} &\object{UGC 08882}  &	13 57 18.7	&54 06 25 &dE,N	&1.0 & 15.28    &    \\
PM&16. & \object{MGC 9-23-21} &\object{MGC 9-23-21}  &	13 57 37.9	&51 58 26 &BCD?	&0.8 & 16.0:{\textsuperscript{2}}&    \\
PM&17. & \object{UGC 08914} &\object{UGC 08914}	 &	13 59 11.9	&52 21 45 &Im	&1.0 & 16.00    &    \\
M &18. & \object{NGC 5474}  &\object{UGC 09013}  &	14 05 02.0	&53 39 44 &Scd	&4.8 & 11.77    &277 \\
M &19. & \object{NGC 5477}  &\object{UGC 09018}  &	14 05 33.1	&54 27 39 &Sm	&1.7 & 14.19    &304 \\
F &20. & \object{DDO 190}   &\object{UGC 09240}  &	14 24 43.4	&44 31 33 &Im	&1.8 & 13.10    &150 \\
M &21. & \object{DDO 194}   &\object{UGC 09405}	 &	14 35 24.6	&57 15 24 &Im	&1.7 & 14.52    &222 \\
\vspace{-2 mm} \\
\hline
\vspace{-2 mm}\\ \small \\ \multicolumn{10}{l}{{\bf Notes:} Columns 5
and 6: 2000.0 epoch coordinates taken from the NED.}\\
\multicolumn{10}{l}{Column 7: Dwarf type reckoned by B.B.~on the
system of Sandage and Binggeli~\cite*{SaB84}}\\
\multicolumn{10}{l}{Columns 8 and 9: Diameter at $\mu = 25 \,
\mathrm{mag}/\sq{\arcsec}$ and total apparent blue magnitude from the
present photometry, from}\\ \multicolumn{10}{l}{ Schmidt \& Boller \cite*{1992AN....313..189S} ({\textsuperscript{1}}), or 
from other sources compiled by one of us (B.\,B.) ({\textsuperscript{2}}). }\\
\multicolumn{10}{l}{Column 10: heliocentric velocity from NED}\\ 
\end{tabular}
\normalsize
\label{sampletable}
\end{table*}

\setcounter{figure}{1}
\begin{figure*}[p]
\vspace{0cm}
\hspace{0cm}
\hbox to \textwidth{\hfil
\vbox to \textheight{\vfil 
\hbox{\resizebox{6.75cm}{!}{\includegraphics{xbscU8215_B.eps.small}}
\resizebox{6.75cm}{!}{\includegraphics{xbscDDO167_B.eps.small}}}
\hbox{\resizebox{6.75cm}{!}{\includegraphics{xbscDDO168_B.eps.small}}
\resizebox{6.75cm}{!}{\includegraphics{UGCA342.eps.small}}}
\hbox{\resizebox{6.75cm}{!}{\includegraphics{xbscDDO169_B.eps.small}}
\resizebox{6.75cm}{!}{\includegraphics{xbscN5204_R_1.eps.small}}}
\vfil}
\hfil}
\caption{$B$-band CCD images (except NGC 5204 for which only a $R$-band
image was available) of the M101 group dwarf members and the dwarfs in
the vicinity of M101, shown in the same order as listed in
Table~\ref{sampletable}.  The scale is the same for all pictures and
is given by the size of one image side $=5\farcm9$. North is up and
east to the left.}
\label{images}
\end{figure*}

\setcounter{figure}{1}
\begin{figure*}[p]
\label{images-b}
\vspace{0cm}
\hspace{0cm}
\hbox to \textwidth{\hfil
\vbox to \textheight{\vfil 
\hbox{\resizebox{6.75cm}{!}{\includegraphics{xbscU8508_B.eps.small}}
\resizebox{6.75cm}{!}{\includegraphics{xbscN5229_B.eps.small}}}
\hbox{\resizebox{6.75cm}{!}{\includegraphics{xbscN5238_B.eps.small}}
\resizebox{6.75cm}{!}{\includegraphics{xbscDDO181_B.eps.small}}}
\hbox{\resizebox{6.75cm}{!}{\includegraphics{xbscU8659_B.eps.small}}
\resizebox{6.75cm}{!}{\includegraphics{xbscDDO183_B.eps.small}}}
\vfil}
\hfil}
\caption{continued}
\end{figure*}

\setcounter{figure}{1}
\begin{figure*}[p]
\vspace{0cm}
\hspace{0cm}
\hbox to \textwidth{\hfil
\vbox to \textheight{\vfil 
\hbox{\resizebox{6.75cm}{!}{\includegraphics{xbscU8833_B.eps.small}}
\resizebox{6.75cm}{!}{\includegraphics{xbscHO4_B.eps.small}}}
\hbox{\resizebox{6.75cm}{!}{\includegraphics{xbscU8882_B.eps.small}}
\resizebox{6.75cm}{!}{\includegraphics{MGC92321.eps.small}}}
\hbox{\resizebox{6.75cm}{!}{\includegraphics{xbscU8914_B.eps.small}}
\resizebox{6.75cm}{!}{\includegraphics{xbscN5474_B.eps.small}}}
\vfil}
\hfil}
\caption{continued}
\label{images-c}
\end{figure*}

\setcounter{figure}{1}
\begin{figure*}[p]
\vspace{0cm}
\hspace{0cm}
\hbox to \textwidth{\hfil
\vbox to \textheight{\vfil 
\hbox{\resizebox{6.75cm}{!}{\includegraphics{xbscN5477_B_1.eps.small}}
\resizebox{6.75cm}{!}{\includegraphics{xbscDDO190_B.eps.small}}}
\hbox{\resizebox{6.75cm}{!}{\includegraphics{xbscDDO194_B.eps.small}}}
\vfil}
\hfil}
\caption{continued}
\label{images-e}
\end{figure*}

\section{Reductions}
\label{reductions}
The images were flat-fielded using combined twilight and dome flats.
The photometry was done with the {\tt MIDAS} package developed by {\tt
ESO}. The images were combined, bias-subtracted and flat\-fielded
using standard procedures. The subsequent reductions were done within
the {\tt SURFPHOT} context in {\tt MIDAS}. The background was
determined by fitting a tilted plane with {\tt FIT/BACKGROUND} and was
checked for correctness by measuring the sky level in different
locations in the field. For each galaxy the centre and the ellipse
parameters (ellipticity, position angle counted coun\-ter-clock\-wise
from the horizontal axis) were determined at the level of
$\sim25^{\mathrm{th}}\mathrm{mag}/\sq{\arcsec}$ by the ellipse fitting
routine {\tt FIT/ELL3} and are given in Table \ref{paandell}.

\begin{table}[!ht]
\caption[]{Parameters of the ellipse fits at approximatively
$25 \mathrm{mag}/\sq{\arcsec}$}
\begin{center}
\begin{tabular}{llllll}
\noalign{\smallskip}
\hline
\noalign{\smallskip}
Number	&Galaxy	& PA [deg] & a[$\arcsec$]& b[$\arcsec$] &$b/a$ \\
\noalign{\smallskip}
\hline
\noalign{\smallskip}
 1.& \object{UGC 08215}		&158& 20.9&15.5&0.74 \\
 2.& \object{DDO 167}		&069& 29.6&17.8&0.60 \\
 3.& \object{DDO 168}		&058&100.&40.2&0.40 \\
 5.& \object{DDO 169}		&044& 63.8&20.6&0.32 \\
 6.& \object{NGC 5204}		&084&137.&88.4&0.64 \\
 7.& \object{UGC 08508}		&027& 54.3&30.2&0.56 \\
 8.& \object{NGC 5229}		&077& 78.0&15.1&0.19 \\
 9.& \object{NGC 5238}		&087& 58.0&39.6&0.68 \\
10.& \object{DDO 181}		&160& 55.7&24.5&0.44 \\
11.& \object{UGC 08659}		&045& 21.3&16.1&0.75 \\
12.& \object{DDO 183}		&121& 67.8&17.8&0.26 \\
13.& \object{UGC 08833}		&052& 26.7&18.9&0.71 \\
14.& \object{HO \,{\sc iv}}	&110&103.&26.8&0.26 \\
15.& \object{UGC 08882}		&163& 28.0&20.6&0.74 \\
16.& \object{MCG 9-23-21}	&   &     &    &     \\
17.& \object{UGC 08914}		&135& 21.5&15.4&0.72 \\
18.& \object{NGC 5474}		&152&103.&99.3&0.96 \\
19.& \object{NGC 5477}		&172& 43.6&30.5&0.70 \\
21.& \object{DDO 190}		&014& 62.3&54.8&0.88 \\
22.& \object{DDO 194}		&050& 42.9&27.6&0.64 \\
\noalign{\smallskip}
\hline
\end{tabular}
\end{center}
\label{paandell}
\end{table}

 These parameters were then used to obtain the total light profile
(growth curve) by integrating the galaxy light in elliptical apertures
of increasing equivalent radius.  A surface brightness profile is 
obtained by differentiating the growth curve. The galaxy profiles
derived in this way include the bright regions that usual ellipse
fitting routines ignore. Circular aperture growth curves were also
obtained as in Bremnes et al.~\cite*{bre98}. The resulting profiles by
the {\em elliptical} aperture photometry are shown in Fig.~\ref{profiles}
and the derived photometric parameters are shown in
Table~\ref{sampletable}. The circular aperture photometry served as a
comparison between the photometry presented here and that of Bremnes
et al.~\cite*{bre98}. This comparison is given in Appendix A, where
it is shown that the agreement between the two methods is excellent.

The profiles are traced down to the level where the errors due to the
fluctuations in the sky level on the profile become dominant.  As
discussed in section \ref{errors}, this represents approx.\
$28.5\,\mathrm{mag}/ \sq{\arcsec}$ in $B$ and $27.5\,\mathrm{mag}/
\sq{\arcsec}$ in $R$.

The photometric calibrations were done using standard methods, with
calibration fields chosen to be relatively close on the sky to the
observed galaxies. These fields were taken from Smith et al.\
\cite*{smith85}. The calibration stars were imaged before and after
imaging every second galaxy.

Galactic absorption values were taken from the NED database, and are
essentially zero for all galaxies in the sample ($A_B = 0.00$ for all
galaxies except NGC\,5204 which has $A_B=0.01$). Therefore no correction
was applied. A correction for internal extinction was not applied either.

\section{Results}
\label{results}
\subsection{Model-free photometric parameters and radial profiles}
\label{photometry}

The global photometric parameters of our objects are listed in Table 
\ref{globalprop}, and the columns represent:\\
\noindent Column 1: number of the galaxy ordered by increasing right ascension\\
Column 2: name of the galaxy\\
Column 3: total apparent magnitude in the $B$ band\\
Column 4: total apparent magnitude in the $R$ band\\
Column 5: effective radius in B $[\arcsec]$\\
Column 6: effective radius in R $[\arcsec]$\\
Column 7: effective surface bright\-ness in B $[\mathrm{mag}/\sq{\arcsec}]$\\
Column 8: effective surface bright\-ness in R $[\mathrm{mag}/\sq{\arcsec}]$\\
Column 9: radius where $\lacute \mu \racute = 25 \, \mathrm{mag}/\sq{\arcsec}$ 
in the $B$ band $[\arcsec]$\\
Column 10: as above, except $\lacute \mu \racute = 26 \, \mathrm{mag}/\sq{\arcsec}$\\
Column 11: as above, except $\lacute \mu \racute = 27 \, \mathrm{mag}/\sq{\arcsec}$\\
Column 12: radius where $\lacute \mu \racute = 25 \, \mathrm{mag}/\sq{\arcsec}$ 
in the $R$ band $[\arcsec]$\\
Column 13: as above, except $\lacute \mu \racute = 26 \, \mathrm{mag}/\sq{\arcsec}$\\
Column 14: as above, except $\lacute \mu \racute = 27 \, \mathrm{mag}/\sq{\arcsec}$\\
Column 15: total $B-R$ $[\mathrm{mag}]$\\

The total apparent magnitude of a galaxy was read off the growth
curve at a sufficiently large radius (i.e. where the growth curve becomes
asymptotically flat). The model-free effective radius was simply read at
half of the total growth curve intensity. The effective surface 
brightness is then given by 
\begin{equation}
\lacute \mu \racute_\mathrm{ eff}[\mathrm{mag}/\sq{\arcsec}]= M + 5
\log(R_\mathrm{ eff}[{\arcsec}]) + 2.
\label{mueffdef}
\end{equation}
All radii refer to {\em equivalent radii}\/ ($r=\sqrt{ab}$, where 
$a$ and $b$ are the major and
minor axis of the galaxy, respectively).

Surface brightness profiles were obtained by differentiating the
growth curves with respect to radius. For 20 of our sample galaxies
the resulting $B$ and $R$ profiles are shown in Fig.~\ref{profiles}.
No profile could be constructed for UGCA 342 due to a bright nearby
star (cf. Sect.~\ref{individual} and Fig.~\ref{images}).  The profiles
have been slightly smoothed (with a running window of width $\approx 3
\arcsec$) and are plotted on a linear radius scale.

\begin{table*}[!ht]
\caption[]{Global photometric properties. See text for explanations.}
\begin{center}
\begin{tabular}{llllllllrrrrrrr}
\noalign{\smallskip}
\hline
\noalign{\smallskip}
Number	&Galaxy	&$B_{\mathrm T}$&$R_{\mathrm T}$&$r^B_\mathrm{ eff}$&$r^R_\mathrm{ eff}$
&$\lacute \mu 
\racute^B_\mathrm{ eff}$&$\lacute \mu \racute^R_\mathrm{ eff}$&$R^B_{25}$&$R^B_{26}$
&$R^B_{27}$&$R^R_{25}$&$R^R_{26}$&$R^R_{27}$&$B-R$\\
(1)&(2)&(3)&(4)&(5)&(6)&(7)&(8)&(9)&(10)&(11)&(12)&(13)&(14)&(15)\\ 
\noalign{\smallskip}
\hline
\noalign{\smallskip}
 1.& \object{UGC 08215}&  16.03 & 15.44& 12.97& 13.27& 23.59&  23.05& 18.8 &25.9  & 32.4   &23.5  &28.3  &      & 0.59\\
 2.& \object{DDO 167} &  15.50 & 14.56& 19.16& 22.76& 23.91&  23.34& 24.4 &33.0  & 43.1   &34.1  &50.9  &62.3  &0.94 \\
 3.& \object{DDO 168} &  13.04 & 12.03& 45.97& 58.55& 23.35&  22.87& 66.8 &92.8  & 130.9  &105.9 &145.1 &184.3 &1.01\\
 5.& \object{DDO 169} &  14.27 & 13.37& 26.18& 31.75& 23.36&  22.87& 38.3 &53.5  & 71.1   &55.1  &68.8  &103.2 &0.90 \\
 6 & \object{NGC 5204}&        & 10.94&      & 43.45&      &  21.13&      &      &        &124.8 &152.3 &178.7 &    \\
 7.& \object{UGC 08508}&  13.88 & 13.07& 23.52& 26.10& 22.74&  22.16& 41.9 &53.7  & 70.5   &54.2  &75.2  &87.0  &0.81 \\
 8.& \object{NGC 5229}&  14.10 & 12.98& 16.65& 18.76& 22.21&  21.35& 35.4 &43.6  & 50.3   &48.2  &60.5  &89.6  &1.12\\
 9.& \object{NGC 5238}&  13.55 & 12.54& 31.20& 38.37& 23.02&  22.46& 49.3 &68.1  & 93.0   &70.8  &115.0 &145.1 &1.01\\
10.& \object{DDO 181}  & 14.36 &      & 30.95&      & 23.81&       & 42.8 &56.0  & 69.2   &      &      &      &    \\
11.& \object{UGC 08659}&  16.16 & 15.41& 16.09& 15.50& 24.20&  23.36& 19.1 &28.0  & 34.0   &25.0  &31.6  &38.2  &0.75\\
12.& \object{DDO 183}  & 14.64 & 13.78& 24.31& 27.47& 23.57&  22.98& 35.6 &45.1  & 55.7   &45.5  &57.4  &70.1  &0.86\\
13.& \object{UGC 08833}&  15.58 & 14.63& 16.73& 19.21& 23.70&  23.05& 23.1 &31.7  & 41.0   &33.0  &42.6  &53.4  &0.95\\
14.& \object{HO \,{\sc iv}}&13.65& 12.58&39.14&44.29& 23.61& 22.82 & 56.3 &71.6  & 93.3   &78.7  &103.9 &124.0 &1.07\\
15.& \object{UGC 08882}&  15.28 & 13.99& 14.73& 15.37& 23.12&  21.92& 24.7 &31.2  & 39.4   &34.2  &44.4  &58.6  &1.29\\
16.& \object{MCG 9-23-21}&     &      &      &      &      &       &      &      &        &      &      &      &    \\
17.& \object{UGC 08914}&  16.00 & 15.19& 16.28& 16.23& 24.06&  23.25& 20.9 &28.5  & 34.3   &27.5  &34.9  &41.2  &0.81\\
18.& \object{NGC 5474}&  11.77 & 10.83& 59.12& 57.10& 22.63&  21.62& 105.0&151.2 &  174.7 &136.3 &169.9 &194.1 &0.94\\
19.& \object{NGC 5477}&  14.19 & 13.16& 24.04& 27.37& 23.10&  22.34&  39.5& 50.9 & 68.1   &51.3  &74.1  &103.2 &1.03\\
21.& \object{DDO 190} &  13.10 & 12.37& 32.19& 35.40& 22.64&  22.11&   59.9& 75.3 &  92.7  &74.6  &92.3  &109.9 &0.73\\
22.& \object{DDO 194} &  14.52 & 13.40& 33.02& 35.98& 24.11&  23.18&   37.4& 56.1 &  73.2  &56.7  &75.0  &93.2 &1.12\\
\noalign{\smallskip}
\hline
\end{tabular}
\end{center}
\label{globalprop}
\end{table*}

\begin{figure*}[!htp]
\hspace{-1cm}\resizebox{\hsize}{!}{\includegraphics{radial_profiles_err.ps.small}}
\caption{Radial surface brightness profiles of the observed dwarf
galaxies in $B$ (lower) and $R$ (upper) except for NGC\,5204 (only
$R$) and DDO\,181 and MCG~9-23-21 for which only $B$ data are
available. The dash-dotted lines represent the exponential fits, as
described in Sect.\ \ref{exponential} and the dashed and dotted lines
represent the error envelopes as described in Sect.\ \ref{errors}. The
radii are all equivalent radii ($r = \sqrt{ab}$). }
\label{profiles}
\end{figure*}

\subsection{The exponential model: fits and parameters}
\label{exponential}
It is well accepted that the radial intensity profiles of dwarf
galaxies can be reasonably well fitted by a simple exponential
\cite{deV59,BC93}.  This applies not only for dwarf ellipticals, but
also for irregulars, if one looks aside from the brighter star-forming
regions and considers the underlying older populations.  These
profiles can be written as
\begin{equation}
I(r) = I_0 \:\exp{\left(-{r\over r_0}\right)}\equiv I_0\:e^{-\alpha r},
\label{exp}
\end{equation}
which in surface bright\-ness representation becomes a straight line
\begin{equation}
\mu (r)= \mu_0 + 1.086\: \alpha r.
\label{expdr}
\end{equation}
The central extrapolated surface bright\-ness $\mu_0$ and the
exponential scale length $1/\alpha$ are the two free parameters of the
exponential fit.  In this work the fits to the profiles were done on
the outer parts of the profiles by a least squares fitting procedure
(note, however, that the very outermost parts were not considered in
the fitting, as they are often ``flaring up'', see below
Sect.~\ref{discussion}).  The best-fitting parameters are listed in
Table~\ref{modelparam}. The best-fitting exponential profiles are
plotted as dash-dotted lines along with the observed profiles in
Fig.~\ref{profiles}.

The deviation from a pure exponential law is expressed by the
difference between the total magnitude of an exponential intensity law
given by
\begin{equation}
M_{\exp}=\mu_0^{\exp} + 5\log\alpha -2.0,
\label{Mexp}
\end{equation}
and the actual measured total magnitude.  The results are shown in
Table~\ref{modelparam}.  The difference is an indication of the
goodness of fit of the exponential intensity profile.  The columns of
Table~\ref{modelparam} are as follows:\\
\noindent Column 1: as column 1 of Table \ref{globalprop}\\
Column 2: as column 2 of Table \ref{globalprop}\\
Column 3: extrapolated central surface bright\-ness according to 
equation \ref{expdr} in $B \,[\mathrm{mag}/\sq{\arcsec}]$\\
Column 4: as above but in $R$\\
Column 5: exponential scale length in $B\,[\arcsec]$\\
Column 6: as above but in $R$\\
Column 7: difference between the total magnitude as derived from the 
exponential model and the true total magnitude in $B$\\
Column 8: as above but in $R$\\
Column 9: radial colour gradient determined from the difference in the 
slopes of the model fits as described in Sect.~\ref{colgradsec} 
$[\mathrm{mag}/\arcsec]$.

\begin{table*}[!ht]
\caption[]{Model parameters. See text for explanations}
\begin{center}
\begin{tabular}{llccccccc}
\noalign{\smallskip}
\hline
\noalign{\smallskip}
Number	&Galaxy	&$(\mu^{\exp}_0)_B$&$(\mu^{\exp}_0)_R$&$1/{\alpha_B}$&$1/{\alpha_R}$
&$M^B_{\exp}-M^B$&$M^R_{\exp}-M^R$&$\left[{\mathrm{ d}(B-R)\over \mathrm{ d}r}\right]_{\exp}$\\ 
(1)&(2)&(3)&(4)&(5)&(6)&(7)&(8)&(9)\\
\noalign{\smallskip}
\hline
\noalign{\smallskip}
 1.& \object{UGC 8215}& 	22.29&21.28 & 7.43 & 6.79& -0.09 & -0.32&	-0.014\\
 2.& \object{DDO 167}& 		22.64&22.24 &10.71 &13.76& -0.01 & -0.01&	 0.022\\
 3.& \object{DDO 168}&  	21.89&21.49 &23.34 &29.59&  0.01 &  0.10&	 0.010\\
 5.& \object{DDO 169}& 		22.15&21.63 &14.74 &17.16&  0.04 &  0.09& 	 0.010\\
 6.& \object{NGC 5204}& 	     &20.29 &      &28.50&       &  0.08&	-0.038\\
 7.& \object{UGC 8508}& 	21.24&20.73 &12.15 &13.89& -0.06 & -0.05&	 0.011\\
 8.& \object{NGC 5229}& 	20.51&20.26 & 8.49 &11.09& -0.23 &  0.05&	 0.030\\
 9.& \object{NGC 5238}& 	21.90&21.27 &17.83 &20.73&  0.09 &  0.15&	 0.009\\
10.& \object{DDO 181}&  	22.00&      &15.02 &     & -0.24 &       & 	      \\
11.& \object{UGC 8659}& 	23.21&22.24 &11.25 & 9.85& -0.21 & -0.14&	-0.014\\
12.& \object{DDO 183}&  	21.89&21.29 &11.82 &13.18& -0.11 & -0.09&	 0.009\\
13.& \object{UGC 8833}& 	22.41&21.82 & 9.60 &11.31& -0.08 & -0.08&	 0.017\\
14.& \object{HO \,{\sc iv}}&    22.11&21.46 &20.68 &24.44& -0.12 & -0.06&      0.008\\
15.& \object{UGC 8882}& 	21.48&20.97 & 7.50 & 9.36& -0.17 &  0.12&	 0.029\\
16.& \object{MCG 9-23-21}&	22.03&      & 6.08 &     &        &       &          \\
17.& \object{UGC 8914}& 	23.02&22.21 &10.71 &10.79& -0.13 & -0.14& 	 0.001\\
18.& \object{NGC 5474}& 	20.75&19.74 &27.85 &26.81& -0.24 & -0.23&	-0.002\\
19.& \object{NGC 5477}& 	21.59&20.99 &12.53 &13.99& -0.09 &  0.10& 	 0.009\\
21.& \object{DDO 190}&  	20.99&20.51 &16.36 &18.10& -0.18 & -0.15& 	 0.006\\
22.& \object{DDO 194}&  	23.03&21.98 &20.25 &20.23& -0.02 &  0.05& 	-0.000\\
\noalign{\smallskip}
\hline
\end{tabular}
\end{center}
\label{modelparam}
\end{table*}

\subsection{Colour gradients}
\label{colgradsec}
\begin{figure*}[!htp]
\hspace{-1cm}\resizebox{\hsize}{!}{\includegraphics{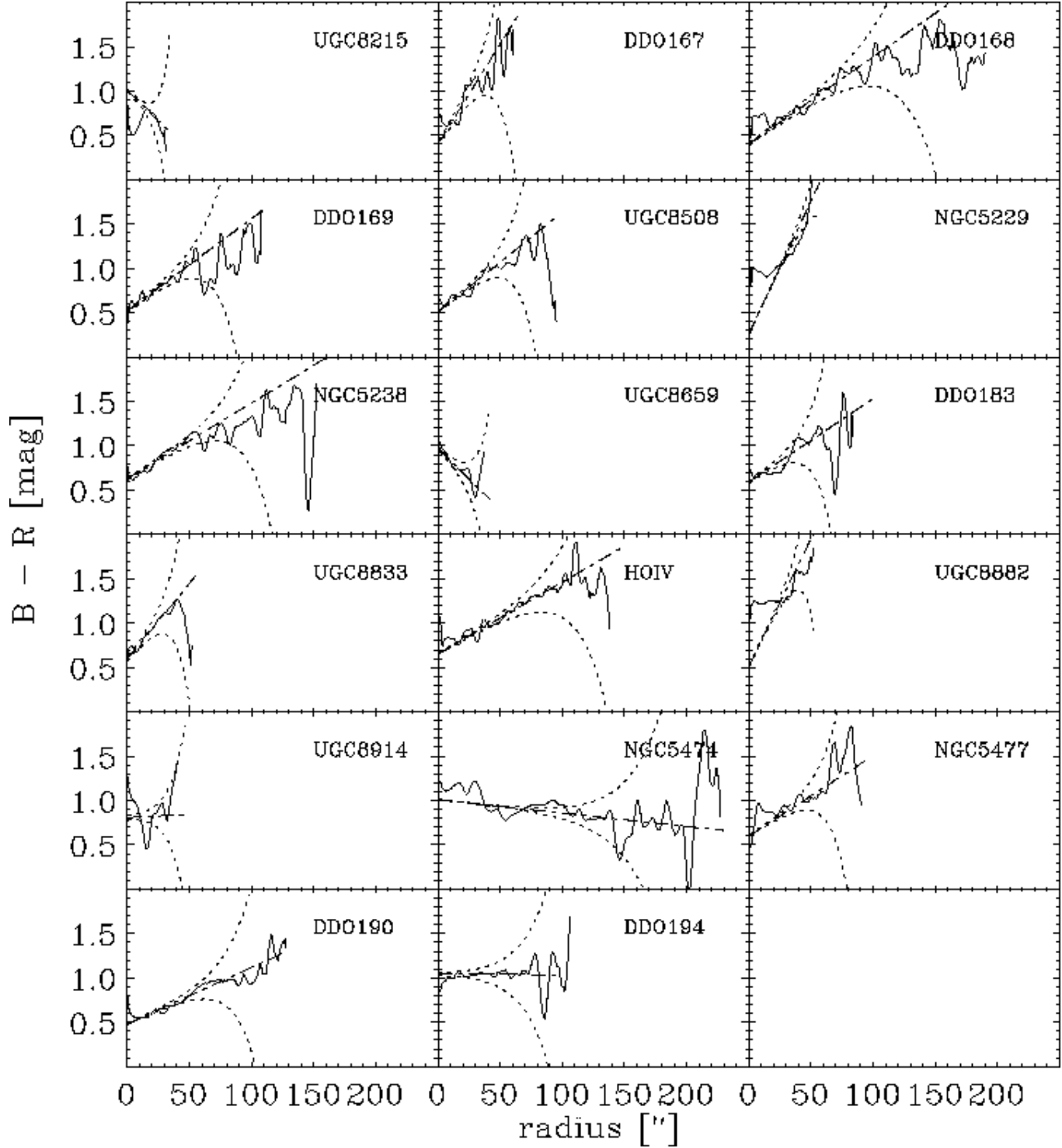}}
\caption{Radial $B-R$ colour profiles. The dot dashed lines represent
the exponential fits, as described in Sect.~\ref{exponential} and the
dotted lines represent the error envelopes as described in 
Sect.~\ref{errors}. }
\label{colourprof}
\end{figure*}

As one can see in Table \ref{modelparam}, the colour gradients of the
galaxies are very small, if not zero ($ \lacute {\mathrm{ d}(B-R)/
\mathrm{ d}r}\racute = 0.012\,\pm 0.013\,\mathrm{mag}/\arcsec$). Many
authors report that colour profiles show very small gradients or are
flat in the case of dwarf galaxies \cite{bre98,patthuan96}. We find
that if the galaxies do show a trend in their colour profiles, they
become slightly redder with increasing radius. The actual colour
profiles together with the difference between the slopes of
exponential fits (dash-dotted) are plotted in Fig.~\ref{colourprof}.

\subsection{Photometric uncertainties}
\label{errors}
Uncertainties in the photometry have multiple sources: calibration
errors, flatfielding and sky subtraction, photon shot noise, readout
noise, contamination by cosmic rays, foreground stars and background
galaxies\ldots

The largest contribution to the uncertainties in the global
photometric parameters is the photometric calibration. As the nights
were non-photometric, one must beware of uncontrollable errors in the
zero-point and the extinction coefficient. The statistical uncertainty
on the photometric calibration is of the order of $0.1$ mag, due to an
uncertainty of $\sim 0.08$ mag and $\sim 0.05$ on the zero-point and
the extinction coefficient, respectively.

The uncertainties on the photometric profiles at low levels are
dominated by the non-flatness of the sky background. The
pixel-to-pixel fluctuations caused by photon shot noise are averaged
out by measuring azimuthally averaged profiles. At typical sky levels
of the order of $\sim 22.7 \,\mathrm{mag}/\sq{\arcsec}$ in $B$, and
$\sim 21.7 \,\mathrm{mag}/\sq{\arcsec}$ in $R$, and a flat-fielding
accurate to $\lesssim 0.5\%$ of the sky background, the sky
fluctuations reach values similar to the the galaxy profiles at
respectively $\sim 28.5$ and $\sim 27.5 \,\mathrm{mag}/\sq{\arcsec}$.

To have a handle on this error along a profile, we have calculated
error envelopes for all our profiles based on their best-fitting
exponentials. The combined uncertainty caused by photon shot noise
from the sky and the galaxy, calculated for azimuthally averaged
$1\arcsec$ annuli, has been added in quadrature with a large-scale sky
flatness and subtraction error term set to a constant $0.5\%$ of the
actual sky electron counts. The error term obtained this way has been
added or subtracted, respectively, from the intensity profiles
corresponding to exponential surface brightness profiles and then
converted to magnitudes to produce the upper and lower error
envelopes. These error envelopes are shown in Fig.~\ref{profiles}
along with the observed and model profiles.  The colour profile error
envelopes, shown in Fig.~\ref{colourprof} as dotted lines, have been
calculated by using the error term as described above for each colour
and applying usual error formulae for logarithms and combining the
errors thus obtained for each colour by quadrature. It is to be noted
that the large scale fluctuation level of $0.5\%$ of the sky
background is an upper limit, most frames showing less variation,
i.e. these error estimates are rather {\em conservative}.  The
calibration zero-point uncertainty is not included in the plots.

\begin{figure}[!htb]
\resizebox{\hsize}{!}{\includegraphics{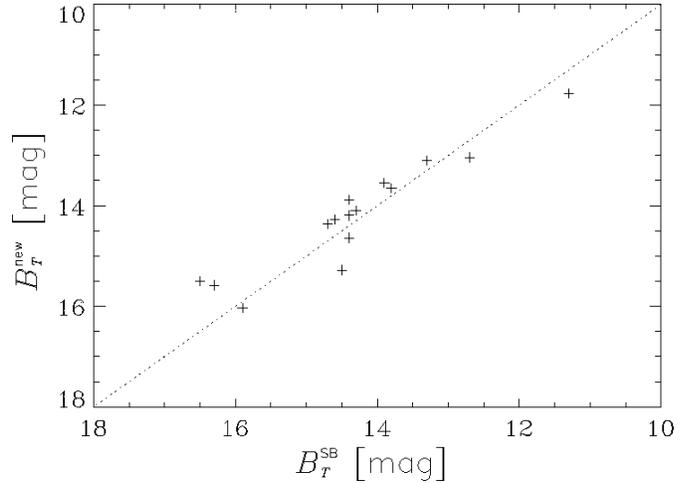}}
\caption{Comparison between the photometry of the present paper and
the data published in Schmidt \& Boller {\protect
\cite*{1992AN....313..189S}}.}
\label{comp_old}
\end{figure}

The errors on the profiles at low luminosity do not influence the
total magnitude to a large extent, but sources projected onto or near
the galaxies do.  We masked out such objects, trying not to eliminate
\ion{H}{ii} regions from the galaxy. An overall assessment of our
photometric accuracy is provided by a comparison with external data.
In Fig. \ref{comp_old} photometry from this paper is compared to data
published in Schmidt \& Boller \cite*{1992AN....313..189S}.  The
agreement is quite good, $\sigma_m \sim 0.13\,\mathrm{mag}$ in the $B$
band.

\section{Discussion}
\label{discussion}
The surface brightness profiles shown in Fig.~\ref{profiles} are quite
typical for dwarf galaxies, being more or less straight lines,
i.e. exponentials, in a large range of intermediate radii, with
deviations from this in the innermost and the outermost parts, i.e. at
small and very large radii; compare, e.g., with the profiles of M81
group dwarfs in Paper I, and of Virgo cluster dwarfs in Binggeli \&
Cameron \cite*{BC93}. The inner deviation from the exponential can be
a luminosity cusp, which is common among dwarf ellipticals, especially
nucleated ones. This feature is well seen in the only dwarf elliptical
of our sample, UGC\,8882. Late-type dwarf galaxies (Sd, Sm, Im), on
the other hand, tend to exhibit a central {\em luminosity deficit}\/
with respect to the best-fitting exponential (see again
Fig.~\ref{profiles}).  In most cases this deficit is simply caused by
the irregular structure of the star-forming region which is confined
to the central part of a galaxy. The peak surface brightness of a
star-forming galaxy can be far off the center as referred to the faint
outer isophotes.  If the profile is centered on the faint outer
isophotes (which in our opinion is the only way a profile makes sense,
as it should, ideally, refer to center of the mass and not the
luminosity distrubution), then the consequence is obviously an
apparent luminosity deficit in the central part. This is dramatically
demonstrated by NGC\,5474, but also Ho\,IV (see Figs.~\ref{images} and
\ref{profiles}). It simply means that the innermost part of the {\em
mean}\/ radial profile of a dwarf irregular (and sometimes even a late
spiral) should not be taken at face value. This is a principal
limitation of one-dimensional ``surface'' photometry of non-symmetric
galaxies.
   
Some of our galaxies show a {\em luminosity excess}\/ above a pure
exponential in the surface brightness profiles at large radii
(Fig.~\ref{profiles}), most strongly in NGC\,5238, but also in
DDO\,168, DDO\,169, DDO\,183, NGC\,5477, DDO\,190, and
DDO\,194. NGC\,5229 also shows an excess in one colour but in this
case our photometric method is clearly not well suited, due to the
galaxy being almost edge-on and warped. On average this excess sets in
at a surface brightness of $\mu \gtrsim 26.7 \pm 0.7
\mathrm{mag}/\sq{\arcsec}$ in $B$ and $\mu \gtrsim 26.1 \pm 0.8\,
\mathrm{mag}/\sq{\arcsec}$ in $R$. This trend was not observed with
our M81 group dwarfs (Paper I), nor was it found in Virgo cluster
irregulars, see Fig.~9 in Binggeli \& Cameron \cite*{BC93}, whose
radial profiles were followed to equally faint levels (most other
photometric studies do not go faint enough, which renders a comparison
difficult).

A possible reason one could think of for this feature is the fact that
in the present photometry the elliptical shape of the running aperture
was fixed for a given galaxy. That ellipse was determined at
approximately $25\,\mathrm{mag}/\sq{\arcsec}$ (cf.\ Sect.\
\ref{reductions}). Suppose that the outermost part of a galaxy, at a
surface brightness level well below $25 \,\mathrm{mag}/\sq{\arcsec}$,
is more spherical than the inner part, or that there is strong
isophotal twisting: then the mean surface brightness profile derived
from apertures of fixed ellipticity will become flatter at large
radii, i.e. will show an excess of the kind discussed here. Galaxies
such as DDO\,168, DDO\,169 and NGC\,5238 show such a behaviour, but at
very low levels. For instance DDO\,168 has an apparent ellipticity of
$\epsilon = 0.60$ and a position angle of $58$ degrees for an ellipse
fit at $\lacute \mu_B \racute = 25 \, \mathrm{mag}/\sq{\arcsec}$ and
at $\lacute \mu_B \racute = 27 \, \mathrm{mag}/\sq{\arcsec}$ these
values are resp.\ $0.42$ and $62$. For DDO\,169 these values are
resp.\ $0.68$, $44$, $0.70$ and $49$ and for NGC\,5238 these values
are resp.\ $0.32$, $87$, $0.33$ and $88$. In the last case $\epsilon$
then rises to $0.41$ in the outermost regions, with unchanged PA. It
seems unlikely that this is the reason for the excess.

A possible physical explanation of the observed break in the surface
brightness profiles -- well-known from the photometry of disk galaxies
-- is of course the presence of two distinct galaxian components (like
bulge and disk).  The existence, in dwarf irregulars, of an 
underlying population of old stars that stretches over a large
characteristic scale length, with a luminous, more concentrated, young
population on top of it, is indeed highly expected.  It has recently
been shown that Local Group dwarf irregulars possess extended old
halos \cite{minniti}.  With this regard it is also interesting that
the observed break in the surface brightness profiles of our dwarfs
seems to be well correlated with a flattening of the corresponding
$B-R$ colour profiles (see Fig.~\ref{colourprof}).  The colour of the
excess light is rather red, with the onset of the flatter part at a
$B-R$ value of $\sim 1.2 \pm 0.1 \,\mathrm{mag}$.

So, have we detected an underlying old halo population on purely
photometric grounds? Unfortunaletly, this cannot be claimed (yet).
The deviations from the exponential luminosity and colour profiles
discussed here are mostly well within the error envelopes, as
indicated in Figs.~\ref{profiles} and \ref{colourprof}. Hence the
significance for a single case is quite weak. However, we note again
that the error envelopes are rather conservative (also see the note on
DDO\,168 below). The tendency seen in so many galaxies is certainly very
suggestive and a follow-up of this question by means of multicolour
photometry of very high accuracy and sensitivity, in order to
definitively prove the reality of the phenomenon, seems very
desirable.

\section{Notes on individual galaxies}
\label{individual}
\noindent{\bf \object{UGC 8215}}:
	small irregular with a relatively steep light
	profile. The central part shows a slight deficit compared to a pure 
	exponential. One of the few galaxies with a slight bluing of its
	colours towards larger radii (but could also have a flat
	colour profile within the errors).

\noindent {\bf \object{DDO 167}}:
	slightly more irregular than UGC 8215, but has identical lightprofile 
	characteristics.

\noindent {\bf \object{DDO 168}}:
	shows an excess in its surface brightness profiles with
	respect to a pure exponential at large radii as well as a
	deficit in the central regions. Its colour profile displays a
	clear flattening from the point at which the luminosity excess
	becomes visible in the profiles. But the error envelopes are
	large at those points both in the radial profiles and colour
	profiles. It is to be noted though that for this galaxy, the
	measured sky counts after subtraction were marginally negative
	around the galaxy. Therefore one should have seen a slight
	deficit in the outer parts of the brightness profiles rather
	than an excess, if the deviations were only due to the
	uncertainties.

\noindent{\bf \object{UGCA 342}}: 
	very irregular, and projected close to a bright star. At a 
	velocity of 388 $\mathrm {km\,s}^{-1}$ it might
	belong to the outskirts of M\,63, which has a redshift of 504 $\mathrm
	{km\,s}^{-1}$. Indeed, UGCA\,342 is only $7.7'$ from M\,63, which is
	less than one optical diameter of M\,63. At a distance of $\approx
	8.5$ Mpc ($H_0\approx 60$) this represents $\approx 19$ kpc in
	projected distance. Some diffuse features are visible in the
	image, especially to the south-east of UGCA\,342, supporting the idea
	that UGCA\,342 is some luminous condensation in the outer part of
	M\,63.    An
	optical image with a radio map of M\,63, provided by Fig.\ 6 in Bosma
	\cite*{bosma}, confirms this idea. Note the similar velocities at 
	the coordinates of UGCA\,342.
	The elongated shape, diffuseness, nearby bright star and
	background sky ``structure'' made accurate photometry impossible.

\noindent {\bf \object{DDO 169}}: 
	like DDO 168, one observes similar
	features in the surface brightness and colour profiles. Extended
	`tail' towards the north. This tail seems to be slightly bluer than
	the rest of the galaxy.

\noindent {\bf \object{NGC 5204}}:
	imaged in $R$ only.

\noindent {\bf \object{UGC 08508}}:
	slight excess in the outer regions as well as a slight deficit in 
	the inner regions. The excess is unlikely to be real. 

\noindent {\bf \object{NGC 5229}}:
	edge on, slightly warped. 

\noindent {\bf \object{NGC 5238}}:
	shows a relatively large excess in the surface brightness
	profiles, as well as a flattening of its colour profile. The 
	characteristics of the excess and the colour profile are similar 
	to those of DDO 168.

\noindent {\bf \object{DDO 181}}:
	imaged only in $B$. 
	Pronounced brightness deficit with respect to a pure 
	exponential in the centre.

\noindent {\bf \object{UGC 08659}}:
	its colour profile shows a bluing with increasing
	radius. Irregular brighter knots.

\noindent {\bf \object{DDO 183}}:
	slight surface brightness excess in the outer parts as well as a 
	deficit in the inner part.
	Possible flattening of the colour profile.

\noindent {\bf \object{UGC 08833}}:
	similar to DDO 167.

\noindent {\bf \object{HO\,{\sc iv}}}: 
	the light is distributed in patches, and the surface brightness 
	profile is almost flat in the inner regions.

\noindent {\bf \object{UGC 08882}}:
	nucleated dwarf elliptical with a flat colour profile and a slightly 
	blue nucleus relative to the bulk of the galaxy.

\noindent {\bf \object{MGC 9-23-21}}: 
	projected very close to a bright star, making accurate
	photometry very difficult. The relatively large light gradient
	around the galaxy more or less excludes accurate photometry of
	the fainter parts, especially considering the small angular
	size.

\noindent {\bf \object{UGC 08914}}: 
	shows a surface brighness profile that falls off faster 
	than an exponential in the outer parts. Flat colour profile 
	except for a red central region relative to the surrounding parts.

\noindent {\bf \object{NGC 5474}}:
	highy asymmetrical galaxy. Too large for the field of view for 
	accurate photometry due to sky subtracion difficulties.  

\noindent {\bf \object{NGC 5477}}:
	also shows a possible surface brightness excess in the outer parts. 
	The colour profile on the other hand looks more usual.

	\noindent {\bf \object{DDO 190}}: 
	redder central part that is 
	slightly less bright than predicted by an exponential profile.

\noindent {\bf \object{DDO 194}}:
    	flat colour profile except for a slightly bluer nucleus.


\section*{Appendix: Elliptical vs.\ circular aperture photometry}
\begin{figure}[!htb]
\resizebox{\hsize}{!}{\includegraphics{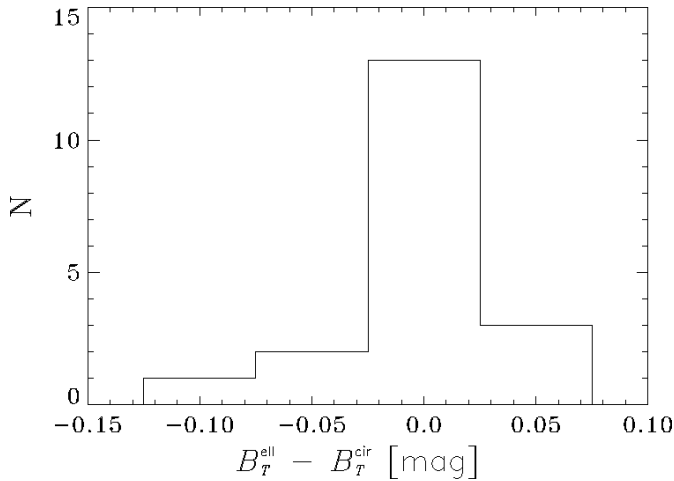}}
\caption{Comparison between the total apparent magnitude $B_T$
obtained with elliptical and circular aperture photometry.}
\label{magbt}
\end{figure}

\begin{figure}[!htb]
\resizebox{\hsize}{!}{\includegraphics{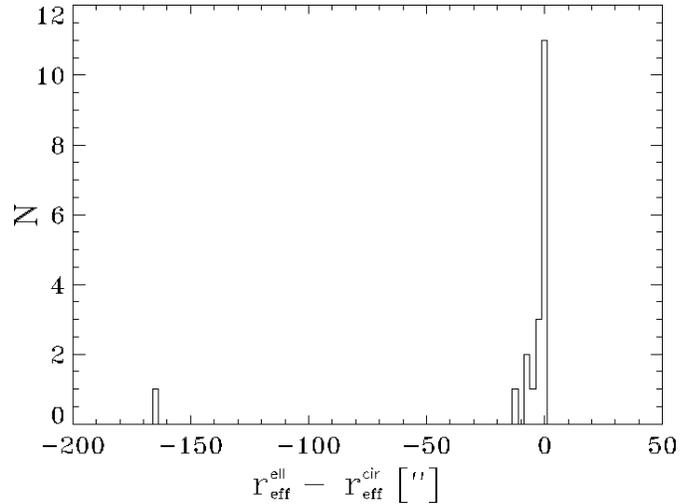}}
\caption{Comparison between the effective radius
obtained with elliptical and circular aperture photometry.
The leftmost object  corresponds to NGC 5229.}
\label{reff}
\end{figure}

\begin{figure}[!htb]
\resizebox{\hsize}{!}{\includegraphics{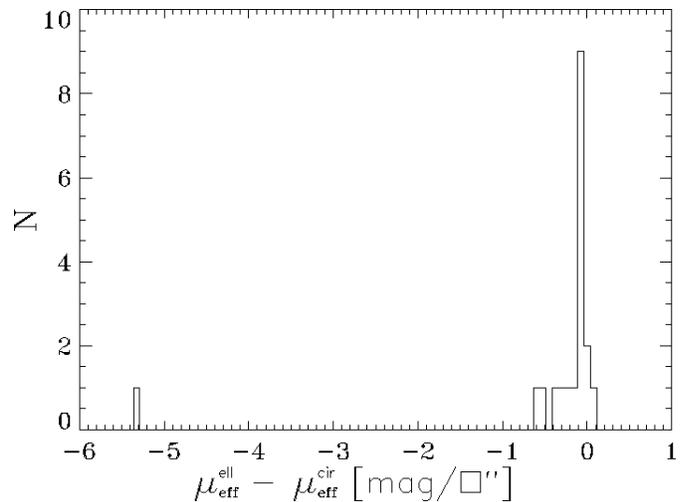}}
\caption{Comparison between the effective surface
brightness obtained with elliptical and circular aperture photometry.
The leftmost object corresponds to NGC 5229}
\label{mueff}
\end{figure}

\begin{figure}[!htb]
\resizebox{\hsize}{!}{\includegraphics{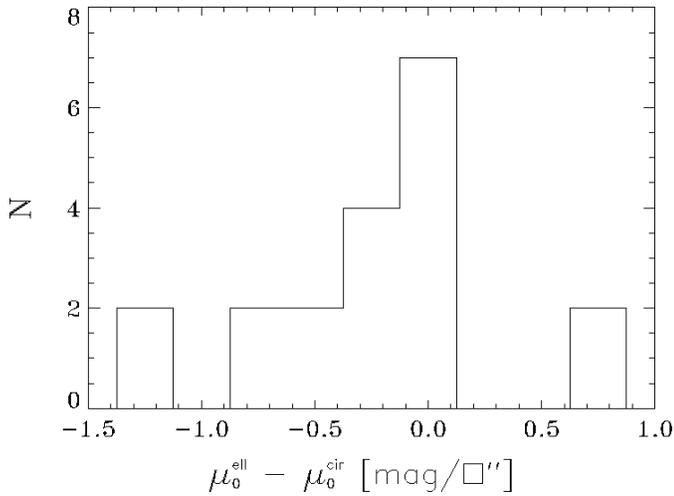}}
\caption{Comparison between the central extrapolated surface
brightness obtained with elliptical and circular aperture photometry. }
\label{s0}
\end{figure}

\begin{figure}[!htb]
\resizebox{\hsize}{!}{\includegraphics{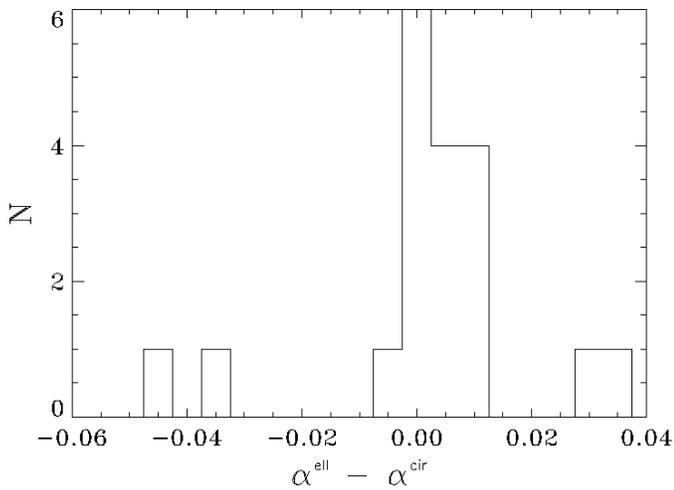}}
\caption{Comparison between the exponential profile parameter $\alpha$
obtained with elliptical and circular aperture photometry. }
\label{alpha}
\end{figure}

We here compare the method used in this paper and that used in Paper
I, i.e. elliptical vs. circular aperture
photometry. Figs. \ref{magbt} to \ref{alpha} compare the different
photometric parameters obtained with the two methods. The comparison
is useful for future studies of a sample of galaxies measured by
either of these two methods.

From Fig. \ref{magbt} one can see the excellent agreement between the
elliptical and the circular aperture photometry as far as
the {\em total} magnitudes are concerned. The vast majority of
the galaxies show differences less than $0.025 \,\mathrm{mag}$.

Fig. \ref{reff}, shows that the effective radius measurements strongly
depend on the apparent shape of the galaxies. Elongated galaxies, as
illustrated by NGC 5229, which has an apparent ellipticity of
$\epsilon=0.81$, show different effective radii as measured by the
different methods. In such cases the circular aperture photometry is
clearly inadequate.

The effective surface brightness obtained by the two methods show
reasonable agreement except, as noted above and for the same reasons,
in the case of very elongated objects, see Fig. \ref{mueff}.

As can be seen in the plots for the exponential fit parameters $\mu_0$
and $\alpha$, Figs. \ref{s0} and \ref{alpha}, the circular apertures
yield slightly larger scale-lengths and lower central
surface brightnesses on average than the elliptical apertures.

\begin{acknowledgements}
T.B.\ and B.B.\ thank the Swiss National Science Foundation for
financial support.  We also thank Frank Thim for taking part in the
observing run and the referee for useful comments.

This research has made use of the NASA/IPAC Extragalactic Database
(NED) which is operated by the Jet Propulsion Laboratory, California
Institute of Technology, under contract with the National Aeronautics
and Space Administration, as well as  NASA's Astrophysics Data System
Abstract Service. 

\end{acknowledgements}

\bibliographystyle{astron}
\bibliography{mnemonic,M101}

\end{document}